\documentclass[12pt,preprint]{aastex}

\slugcomment{To appear in ApJ}

\shorttitle{Constraining the Physical Properties of the Near-Earth
  Object 2009~BD} \shortauthors{Mommert et al.}

\begin{document}


\title{Constraining the Physical Properties of Near-Earth
  Object 2009~BD}


\author{M. Mommert}
\affil{Department of Physics and Astronomy, Northern Arizona
  University, PO Box 6010, Flagstaff, AZ 86011, USA}

\email{michael.mommert@nau.edu}

\author{J.~L. Hora}
\affil{Harvard-Smithsonian Center for Astrophysics, 60 Garden Street,
  MS 65, Cambridge, MA 02138-1516, USA}

\author{D. Farnocchia}
\affil{Jet Propulsion Laboratory, California Institute of Technology,
  Pasadena, CA 91109, USA}

\author{S.~R. Chesley} \affil{Jet Propulsion Laboratory, California
  Institute of Technology, Pasadena, CA 91109, USA}

\author{D. Vokrouhlick\'{y}} \affil{Institute of Astronomy, Charles
  University, V Hole\u{s}ovi\u{c}k\'{a}ch 2, CZ-18000, Prague 8, Czech
  Republic}

\author{D.~E. Trilling}
\affil{Department of Physics and Astronomy, Northern Arizona
  University, PO Box 6010, Flagstaff, AZ 86011, USA}

\author{M. Mueller} 
\affil{SRON Netherlands Institute for Space
  Research, Postbus 800, 9700 AV, Groningen, The Netherlands}

\author{A.~W. Harris}
\affil{DLR Institute of Planetary Research, Rutherfordstrasse 2, 12489
  Berlin, Germany}

\author{H.~A. Smith}
\affil{Harvard-Smithsonian Center for Astrophysics, 60 Garden Street,
  MS 65, Cambridge, MA 02138-1516, USA}

\and

\author{G.~G. Fazio}
\affil{Harvard-Smithsonian Center for Astrophysics, 60 Garden Street,
  MS 65, Cambridge, MA 02138-1516, USA}




\begin{abstract}
  We report on {\it Spitzer Space Telescope} IRAC observations of 
  near-Earth object (NEO) 2009~BD that were carried out in support of
  the NASA Asteroid Robotic Retrieval Mission (ARRM) concept. We did
  not detect 2009~BD in 25~hrs of integration at 4.5~$\mu$m. Based on
  an upper-limit flux density determination from our data, we present
  a probabilistic derivation of the physical properties of this
  object. The analysis is based on the combination of a thermophysical
  model with an orbital model accounting for the non-gravitational
  forces acting upon the body. We find two physically possible
  solutions.  The first solution shows 2009~BD as a $2.9\pm0.3$~m
  diameter rocky body ($\rho=2.9\pm0.5$~g~cm$^{-3}$) with an extremely
  high albedo of $0.85_{-0.10}^{+0.20}$ that is covered with
  regolith-like material, causing it to exhibit a low thermal inertia
  ($\Gamma=30_{-10}^{+20}$~SI units). The second solution suggests
  2009~BD to be a $4\pm1$~m diameter asteroid with
  $p_V=0.45_{-0.15}^{+0.35}$ that consists of a collection of
  individual bare rock slabs ($\Gamma = 2000\pm1000$~SI units, $\rho =
  1.7_{-0.4}^{+0.7}$~g~cm$^{-3}$). We are unable to rule out either
  solution based on physical reasoning. 2009~BD is the smallest
  asteroid for which physical properties have been constrained, in
  this case using an indirect method and based on a detection limit,
  providing unique information on the physical properties of objects
  in the size range smaller than 10~m.
\end{abstract}


\keywords{minor planets, asteroids: individual (2009 BD) --- infrared: planetary systems}



\section{Introduction}
\label{lbl:introduction}


The physical properties of near-Earth objects (NEOs) provide important
hints on their origin, as well as their past physical and orbital
evolution. The most accessible physical properties are the diameter,
$d$, and the geometric albedo, $p_V$, which have been measured for
more than 1000 NEOs with diameters down to slightly less than 100~m in two
large-scale programs, the Warm {\it Spitzer} NEO survey ``ExploreNEOs''
\citep{Trilling2010}, and the ``NEOWISE'' project \citep{Mainzer2011},
using the {\it Wide-field Infrared Survey Explorer}
\citep[{\it WISE},][]{Wright2010}. Recently, \citet{Mainzer2013} measured
the sizes and albedos of the smallest optically discovered NEOs ($d >
10$~m) from NEOWISE data. Little is known about the physical
properties of even smaller NEOs, which constitute the bulk of the NEO
population. Knowledge of the physical properties of such small NEOs,
some of which pose an impact threat to the Earth, is of importance for
understanding their evolution and estimating the potential of
destruction in case of an impact, as well as for designing the most
promising mitigation mission.

Further information on asteroid physical properties are available only
for select objects with relatively large sizes, which make up only a
fraction of the whole asteroid population. Such properties include,
but are not limited to, the bulk density, $\rho$, thermal inertia,
$\Gamma$, and the obliquity, $\gamma$, all of which affect
non-gravitational forces that act upon the body and alter its orbit
compared to a Keplerian one. Two important effects are the Yarkovsky
effect \citep[see, e.g.,][]{Bottke2006} and the solar radiation pressure
\citep{Vokrouhlicky2000}.

The bulk density, $\rho$, provides the simplest way of gaining insight
into asteroid interiors. Solid rock, or monolithic, bodies have high
bulk densities ($\rho\sim3$~g~cm$^{-3}$), whereas those of
rubble-pile bodies, aggregates of smaller particles that are
consolidated by their self-gravity or other adhesive forces
\citep{Chapman1978}, can be significantly lower as a result of
``macroporosity.'' Macroporosity refers to cavities and void spaces
that occur between the irregularly shaped individual constituents
\citep[see][for a discussion]{Richardson2002,
  Britt2002}. \citet{Britt2002} found that most asteroids show a
significant degree of macroporosity, in support of the hypothesis
that most asteroids must have been disrupted in the course of
high-velocity impacts over the age of the Solar System
\citep{Chapman1978}. Small asteroids are generally thought of as being
individual pieces of compact debris that were generated in disruptive
collisions \citep{Pravec2002}; hence, their macroporosity is expected
to be low and their bulk density high compared to that of rubble pile
asteroids.

Thermal inertia, $\Gamma$, describes the ability of the surface
material to store thermal energy: high-thermal-inertia material heats
up slowly and re-emits the thermal energy only gradually, whereas
low-thermal-inertia material can be approximated as being in
instantaneous thermal equilibrium with the incoming insolation
\citep[see, e.g.,][]{Spencer1989}. Examples for materials of low and
high thermal inertia are regolith \citep[30--50~SI units,][1~SI unit
equals 1~J~m$^{-2}$~s$^{-0.5}$~K$^{-1}$]{Spencer1989,Putzig2005} and
bare rock \citep[${>}2500$~SI units,][]{Jakosky1986},
respectively. Measurements of the thermal inertia of medium-to-large
sized NEOs ($d > 100$~m) revealed values of 100--1000~SI units
\citep{Delbo2007}.

Both the thermal inertia and the bulk density of asteroids can be
derived by modeling the effect of non-gravitational perturbations on
the object's orbit \citep[see, e.g.,][]{Chesley2014}. Assuming a
homogeneous bulk density of the constituent particles, usually derived
from laboratory measurements of meteorite equivalent material, allows
for constraining the degree of macroporosity of the asteroid.

NEO 2009~BD was discovered on January 16, 2009, at a distance from the
Earth of only 0.008~au \citep{Buzzi2009}. Its orbit is very
Earth-like with a period of 400~days (JPL Solution 41). The escape
velocity of 2009~BD with respect to the Earth is among the lowest
for known objects ($v_{\infty} \sim 1$~km~s$^{-1}$), making it
a worthwhile candidate mission target. 

2009~BD is considered the primary candidate mission target for NASA's
Asteroid Robotic Retrieval Mission
\citep[ARRM,][]{NASAAsteroidInitiativeWebsite}. The mission concept
involves capturing an asteroid and dragging it onto a new trajectory
that traps it in the Earth-Moon system, where it will be further
investigated by astronauts. As a result of 2009~BD's Earth-like orbit,
its next encounter with the Earth-Moon system will be in late 2022,
when the proposed capture through ARRM would take place. The current
mission design requires the target asteroid to have a diameter of
7--10~m and a total mass of ${\sim}500$~metric tons
\citep{NASAmissionwebsite}. The orbital parameters and absolute
magnitude, $H$, which is the apparent magnitude of an object at a
distance of 1~au to the Sun and the observer, of 2009~BD are
well-known, providing accurate orbital predictions
\citep{Micheli2012}. However, there is no albedo-independent
determination of its diameter, which is a crucial variable in the ARRM
mission planning.

We report here in observations of 2009~BD using the IRAC camera on the
{\it Spitzer Space Telescope}, which provides the only practical means
to constrain the physical properties of 2009~BD in the next
decade. The main goals of our observations were two-fold: measure the
size and therefore determine the suitability of 2009~BD as an ARRM
mission target, and constrain other physical properties like bulk
density and thermal inertia of an asteroid at a size range that is so
far unprecedented.

\section{Observations}
\label{lbl:observations}

We observed 2009~BD with the Infrared Array Camera
\citep[IRAC,][]{Fazio2004} on-board the {\it Spitzer ​Space Telescope}
\citep{Werner2004} in Program ID 90256 using Director's Discretionary
Time. A total of 25~hrs of observation time was split into three
Astronomical Observation ​ Requests (AORs): 49092096 (observation
mid-time: Oct 13, 2013, 16:23 UTC; 8~hrs elapsed time), 49091840 (Oct
14, ​00:20 UTC; 8~hrs), and ​ 49091584 (Oct 14, 20:54 UTC; 9~hrs). The
observation window was selected based on {\it Spitzer} observability.
Based on flux density predictions derived with the near-Earth asteroid
thermal model \citep[NEATM,][]{Harris1998}, a detection in IRAC
channel 1 (3.6~$\mu$m) seemed to be unlikely. Hence, all available
observing time was used on channel 2 (4.5~$\mu$m) observations, where
the predicted flux density was greater than the predicted $5\sigma$ IRAC channel
2 sensitivity during the observation window.

In our observations, individual AORs used the ``Moving Single'' object
mode to track in the moving frame of 2009~BD. A medium cycling dither
pattern was used with a 100~sec frame time.
In order to provide the most ​
accurate pointing during our observations, the JPL Horizons online
Solar System data and ephemeris computation routine, which provides
the {\it Spitzer} pointing information, was updated to include
non-gravitational effects in the prediction of the orbit. We modeled
non-gravitational perturbations as
\begin{equation}
\mathbf a_{NG} = (A_1 \hat{\mathbf r} + A_2 \hat{\mathbf t})
\left(\frac{1 \ \mathrm{au}}{r}\right)^2\ ,
\end{equation}
where $\hat{\mathbf r}$ and $\hat{\mathbf t}$ are the radial and
transverse directions, respectively, and $r$ is the heliocentric
distance. $A_2/r^2$ models the transverse component of the Yarkovsky
effect \citep{Bottke2006}, whereas $A_1/r^2$ models the solar
radiation pressure \citep{Vokrouhlicky2000} and the radial component
of the Yarkovsky effect. This is similar to the comet-like model for
non-gravitational perturbations \citep{Marsden1973}.  The orbital fit
(JPL Solution 41) to the observations yields $A_1 = (57.03 \pm 7.79)
\times 10^{-12}$ au/d$^2$ and $A_2 = (-113.02 \pm 7.89) \times
10^{-14}$ au/d$^2$ \citep[see also the entry for 2009~BD in the][as of
October 24, 2013]{SBDB}. The correlation coefficient between $A_1$ and
$A_2$ is 0.81. The orbital fit is based on 180 optical observations
over the interval from 2009-Jan-16.3 to 2011-Jun-21.0. The positional
uncertainty of 2009~BD as seen by {\it Spitzer} at the time of the
observations was​ ${\pm5}$\farcs0 in right ascension and
${\pm}$0\farcs4 in declination at a $3\sigma$ confidence level. For
comparison, IRAC offers a square field of view with a width of
5\farcm2\ and a pixel scale of 1\farcs2/pixel \citep{WSOM}.
Hence, the accuracy of the orbit determination for 2009~BD is
sufficient to determine its position to within a few IRAC pixels (see
Section \ref{lbl:discussion_observations} for a more detailed
discussion).

The data were reduced using a method tailored to faint NEOs, based on
the ExploreNEOs program \citep{Trilling2010}. In this method, a mosaic
of the field is​ constructed from the dataset itself and then
subtracted from the individual Basic Calibrated Data (BCD)
frames. During these observations, the target had an apparent motion
of ${\sim}$0\farcs4 during each 100~sec frame, so background stars
were trailed only very slightly in individual BCDs. ​We were therefore​
able to generate a high signal-to-noise mosaic of the field to
subtract from the BCDs. After subtraction of the background mosaic,
regions near the peaks of background sources (which had small
residuals) and bright cosmic ray artifacts were also masked in the
individual BCDs, in order to minimize the background noise. The
processed BCDs were then mosaicked in the reference frame of the
moving object for each AOR, and the results from the three AORs
combined to produce a final mosaic that included the full set of 800
100-second frames.

\begin{figure}
\epsscale{.80}
\plotone{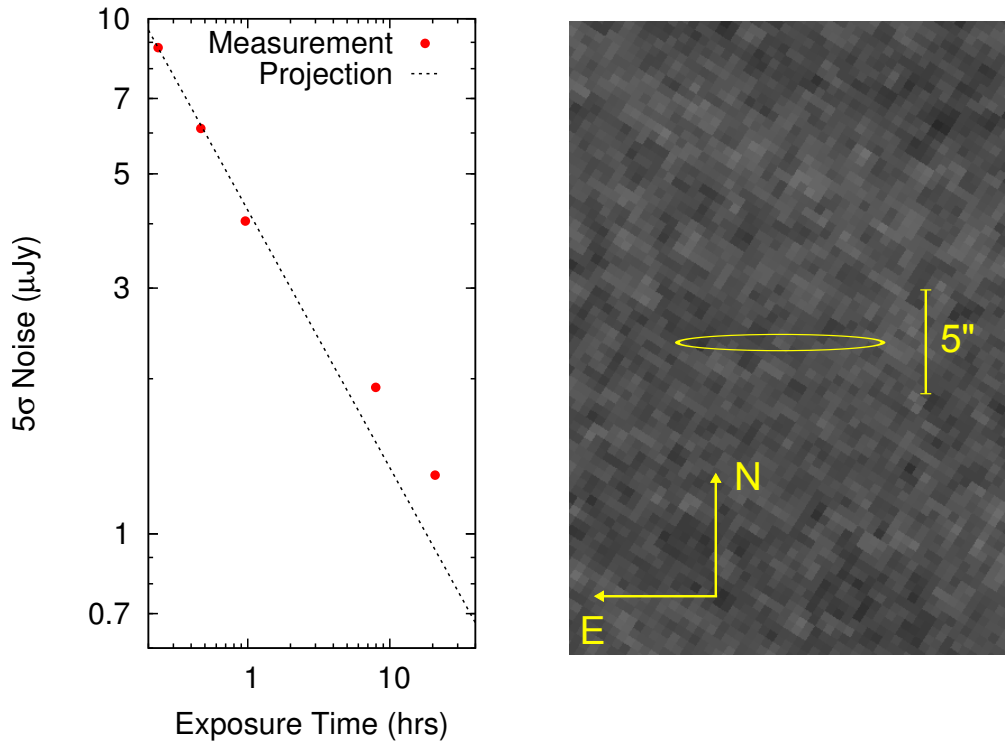}
\caption{{\bf Left:} IRAC channel 2 noise level prediction and
  measurements. Measured sensitivities are derived from ExploreNEOs
  data \citep{Trilling2010, Trilling2013}. The projection is based on
  a $1/\sqrt{t}$ relation, where $t$ is the integration time. The
  predicted sensitivity for our 25~hrs observation was 0.78~$\mu$Jy at
  a $5\sigma$ level. The two data-points with the longest integration
  times have been measured as part of this work. {\bf Right:} Excerpt
  from the final co-added map derived from our observations. The
  predicted position of 2009~BD is indicated with an ellipse,
  indicating the $3\sigma$ uncertainty interval in right ascension and
  declination. The measured noise level is 0.78~$\mu$Jy at a $3\sigma$
  level, which is higher than predicted (see left panel and the
  discussion in Section
  \ref{lbl:discussion_observations}).\label{fig:sensitivity_finalimage}}
\end{figure}

We did not detect 2009~BD in this final co-added map (Figure
\ref{fig:sensitivity_finalimage}, right), from which we derive a
$3\sigma$ upper limit to the flux density of 2009~BD of 0.78~$\mu$Jy.

\section{Modeling Method}
\label{lbl:modeling}

The lack of a clear detection of 2009~BD in our observations precludes
a direct determination of its physical properties. In order to be able
to indirectly constrain the physical properties of 2009~BD, we take a
probabilistic approach that combines a thermophysical model with a
model of the non-gravitational effects on the asteroid's orbit. Based
on the upper-limit flux density provided by our {\it Spitzer}
observation and available astrometric measurements, the combination of
both models allows us to constrain the physical properties of
2009~BD. The combination of the two models provides proper accounting
for the mutual dependencies of the individual physical properties that
impact both models, which would not be possible using a more simple
thermal model.

We model the non-gravitational effects on the orbit of 2009~BD, namely
the solar radiation pressure and the Yarkovsky effect, as a function
of $d$, $\Gamma$, $\rho$, $\gamma$, and other parameters in a numerical
approach. For the solar radiation pressure we assume
\citep{Vokrouhlicky2000}
\begin{equation} {\mathbf a}_{\mathrm{SRP}} = \left (1 + \frac{4}{9}\
    A \right ) \cdot \Psi \cdot \frac{G_S}{c} \cdot
  \frac{\hat{\mathbf r}}{r^2},
  \label{eqn:srp}
\end{equation}
where $A$ is the Bond albedo, $\Psi$ is the area-to-mass ratio of the
object \citep[see, e.g.,][]{Micheli2012}, $G_S$ = 1370 W/m$^2$ is the
solar constant, and $c$ is the speed of light. For the Yarkovsky
effect, we use the model approach described by
\citet{Vokrouhlicky2000b}, which fully captures both the diurnal and
the seasonal components of the Yarkovsky effect. The model asteroid is
assumed to be spherical and the heat transfer is solved analytically
using the linearized heat transfer equation \citep{Vokrouhlicky1998,
  Vokrouhlicky1999}. By fitting all available astrometric data of
2009~BD, the model derives $\rho$ and $\Gamma$ as a function of
$\gamma$ and $d$, as well as the goodness-of-fit parameter $\chi^2$.

The thermophysical model approximates the surface temperature
distribution of 2009~BD and is used in this work to determine the
thermal-infrared emission from its surface as a function of its
physical properties. The model accounts for the spin axis orientation
(represented by $\gamma$), rotational period, thermal inertia, and
surface roughness. We assume a spherical shape of 2009~BD; the
diameter derived with the model is hence the diameter of a sphere with
the same volume as the real shape of 2009~BD. Surface roughness causes
infrared beaming, an effect that focuses thermal emission radiated
towards the observer, and is modeled as emission from spherical
craters \citep[see][for more details]{Mueller2007}. The model, which
is mostly identical to the one discussed by \citet{Mueller2007},
solves the heat transfer equation numerically for a large number of
plane surface facets that form a sphere. The monochromatic flux
density derived by the model is turned into an IRAC channel 2 in-band
flux density, i.e., it is color corrected, using the appropriate
channel 2 response function and assuming a black-body spectrum of the
instantaneous thermal equilibrium temperature for 2009~BD
\citep{Trilling2010}. Furthermore, the contribution from reflected
solar light is added to the calculated flux density using the method
described by \citet{Mueller2011} (and references therein), assuming an
infrared/optical reflectance ratio of 1.4.

In both the orbital and the thermophysical model we adopt the absolute
magnitude $H = 28.43 \pm 0.12$ \citep{Micheli2012}, the photometric
slope parameter $G = 0.18 \pm 0.13$ \citep[derived as the average from
all $G$ measurements of asteroids, see][]{SBDSE}, and the rotation
period $P = 2^{(2\pm0.5)}$~hrs \citep[which is consistent with
observations by][$P \geq 3$~hrs]{Tholen2013} throughout this work.

\section{Results}
\label{lbl:results}

\begin{figure}
\epsscale{.80}
\plotone{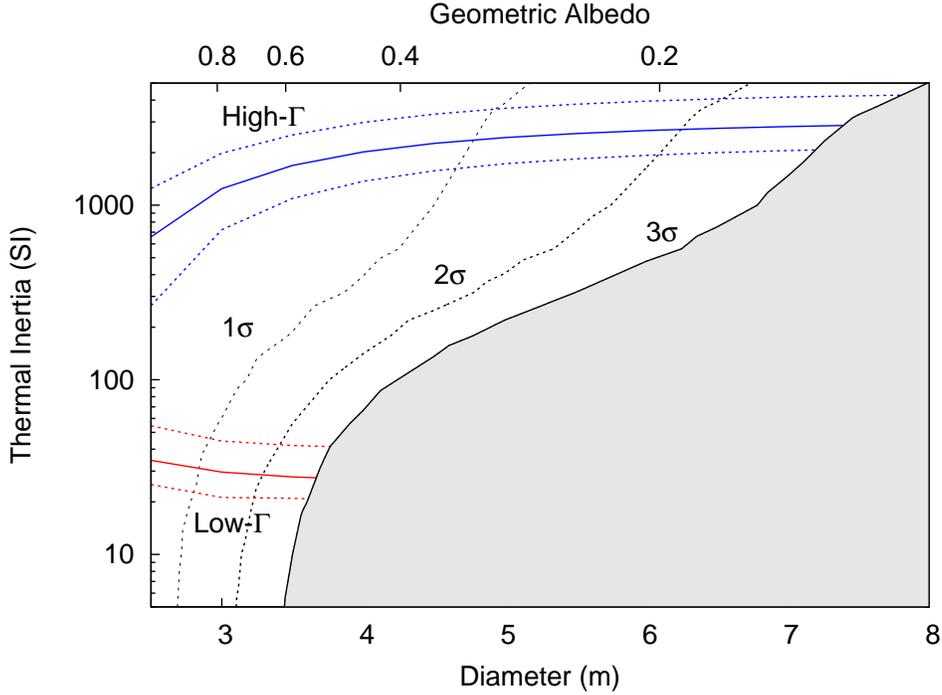}
\caption{Flux density distribution of 2009~BD in diameter-thermal
  inertia space, according to the thermophysical model. The black
  [dotted, dashed, solid] lines illustrate the curvature of the area
  in which 2009~BD would have a flux density that is equal to our
  derived [1, 2, 3]$\sigma$ upper-limit flux density. Grey areas have
  flux densities higher than our derived $3\sigma$ upper-limit flux
  density; there is a 99.7\% probability that 2009~BD must be located
  to the left of the black solid line. This plot is based on the
  assumption that 2009~BD has a smooth surface, spins rapidly
  ($P=3$~h), and has $\gamma=180\degr$; this configuration provides
  the lowest possible flux densities, and hence the largest possible
  range in diameter for 2009~BD. For different configurations, the
  black lines are shifted to smaller diameters. The red and blue lines
  represent the two possible solutions of the orbital model in thermal
  inertia for the possible ranges in diameter (dashed lines illustrate
  $1\sigma$ uncertainties, see text). The flux densities used in the
  production of this plot represent IRAC channel 2 in-band flux
  densities and include contributions from reflected solar
  light.\label{fig:d_TI_map}}
\end{figure}

The mutual dependencies among physical properties used by the orbital
and the thermophysical model require an iterative solution of the
problem. In a first approximation, we constrain the possible ranges of
$\gamma$ and $d$. As the negative value of $A_2$ suggests a retrograde
rotation \citep[see][]{Farnocchia2013}, we sample the obliquity
$\gamma$ from $90\degr$ to $180\degr$. We investigate the possible
range of $d$ using the thermophysical model, based on an $3\sigma$
upper-limit flux density measurement (0.78~$\mu$Jy). Figure
\ref{fig:d_TI_map} shows the predicted flux density at 4.5~$\mu$m as a
function of the diameter and the thermal inertia for the faintest
possible model asteroid, providing the largest possible diameter range
for 2009~BD with a smooth surface, $\gamma = 180\degr$, and the
shortest rotation period consistent with observations
\citep[$P\geq3$~hrs,][]{Tholen2013}. From this plot, we constrain the
possible diameter range of 2009~BD to ${<}8$~m, as a result of our
upper-limit flux density determination. Note that we do not force a
lower-limit diameter constraint, so we do not {\it apriori} exclude high
geometric albedos.

Sampling $d<8$~m and $90\degr < \gamma < 180\degr$ with the orbital
model provides further constraints on 2009~BD's
properties. Intriguingly, we find for each pair ($d$, $\gamma$) two
local minima in the orbital fit $\chi^2$, representing two physically
possible solutions. The ``low-$\Gamma$'' solution displays a low
thermal inertia of the order of 10~SI units with a high bulk density
$\rho$, whereas the ``high-$\Gamma$'' solution stands out with a
thermal inertia of more than 1000~SI units and a low bulk
density. Figure \ref{fig:d_TI_map} shows both solutions in thermal
inertia as a function of the diameter. Based on the orbital fit
solutions, we can also further constrain the obliquity (low-$\Gamma$:
$\gamma=170\degr_{-20}^{+10}$, high-$\Gamma$: $\gamma =
180\degr_{-5}^{+0}$, uncertainties are $1\sigma$) and we can confidently
rule out that 2009~BD is smaller than 2.6~m. For diameters smaller than
that, the orbital model is unable to converge on a physically
meaningful solution (see Section \ref{lbl:discussion_properties} for a
detailed discussion).

\begin{figure}
\epsscale{.80}
\plotone{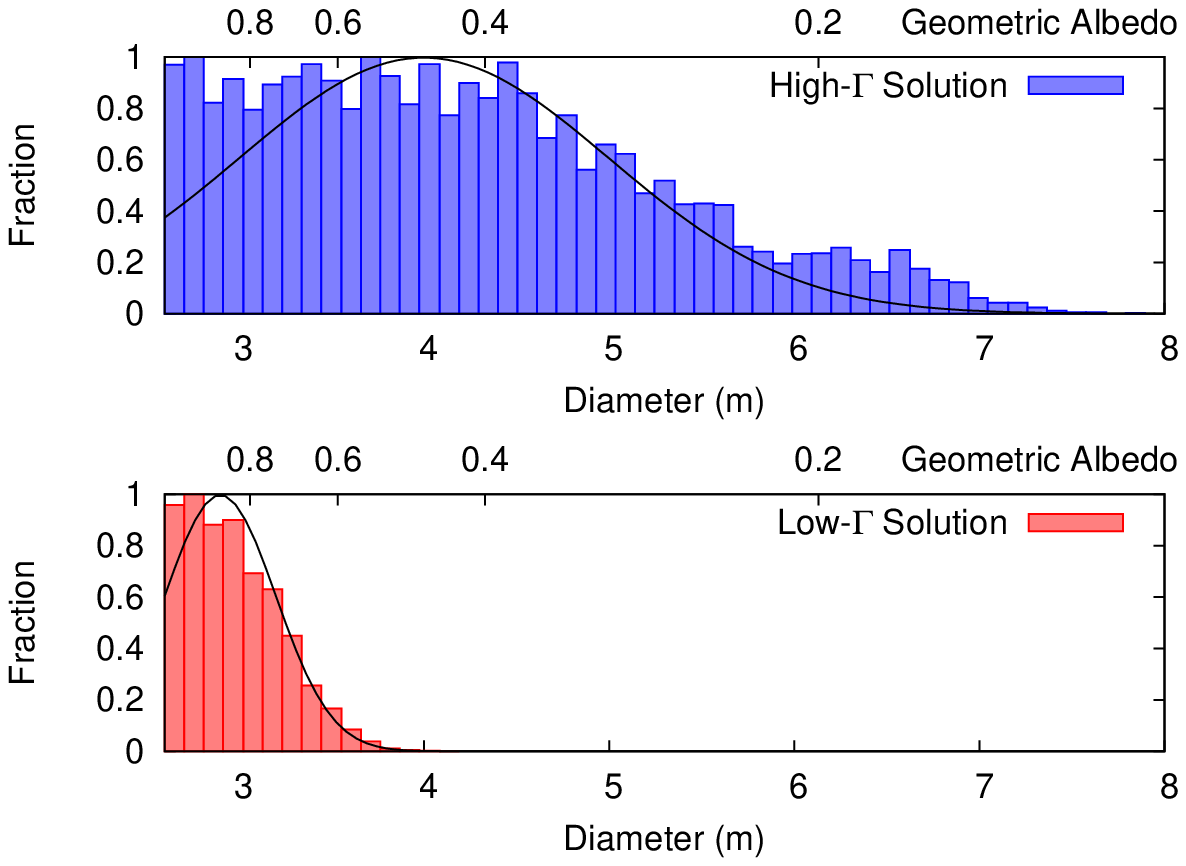}
\caption{Diameter distributions of the low-$\Gamma$ (bottom) and
  high-$\Gamma$ (top) solutions. Over-plotted black lines are normal
  distributions fitted to the distributions. We derive $2.9\pm0.3$~m
  for the low-$\Gamma$ solution and $4\pm1$~m for the high-$\Gamma$
  solution with a lower limit of 2.6~m for both
  solutions.\label{fig:diameter_distributions}}
\end{figure}

We utilize our intermediate results to derive diameter distributions
for both solutions using the thermophysical model, based on the
thermal inertia constraints and the {\it Spitzer} upper-limit flux
density measurement. We generate a sample of synthetic objects with
pairs ($d$, $\Gamma$) that comply with normal distributions around the
thermal inertia solutions shown in Figure \ref{fig:d_TI_map}. We
sample the other model input parameters ($H$, $G$) according to normal
distributions or log-normal distributions ($P$) within the ranges
given in Section \ref{lbl:modeling}. We use
$\gamma=170\degr_{-20}^{+10}$ and $\gamma = 180\degr_{-5}^{+0}$ for
the low-$\Gamma$ and the high-$\Gamma$ solution, respectively, and for
the surface roughness we randomly pick one of four different roughness
models \citep[no, low, default, and high roughness;
see][]{Mueller2007}. We model each synthetic sample object and derive
its IRAC in-band flux density combined with contributions from
reflected solar light, which we then compare with the $3\sigma$
upper-limit flux density as derived from our observations. In case the
sample object flux density is lower than the upper limit, we regard
this individual synthetic object a possible configuration for 2009~BD
and add its diameter to the distribution. The final solution-specific
diameter distributions are shown in Figure
\ref{fig:diameter_distributions} and the derived nominal values and
uncertainties are listed in Table \ref{tbl:solutions}. Nominal values
represent the median values of the respective distributions;
uncertainties are standard deviations, $\sigma$, of a normal
distribution fitted to those values higher than the median of the
distribution. Figure \ref{fig:diameter_distributions} shows that this
approach reasonably describes the range of values lower than the
median, which deviates from the shape of a normal distribution. We
derive albedo values for the high and low-$\Gamma$ solutions
similarly, using separate values of $\sigma$ for albedos higher and
lower than the median, allowing for asymmetric uncertainties.

\begin{figure}
\epsscale{.80}
\plotone{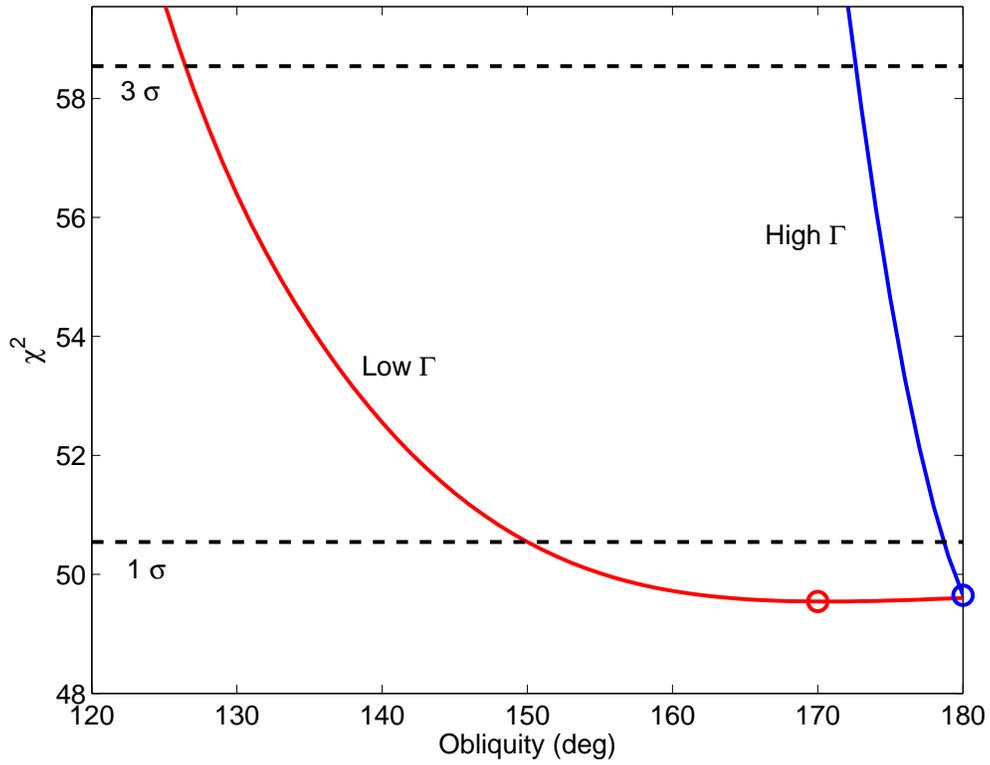}
\caption{$\chi^2$ of the orbital fit as a function of the obliquity.
  The circles indicate the minima of the solutions (low and
  high-$\Gamma$). The low-$\Gamma$ solution allows for obliquities
  $\gamma \geq 150\degr$ at the $1\sigma$ level, $\gamma \geq
  130\degr$ at the $3\sigma$ level. The high-$\Gamma$ solution allows
  for $\gamma > 175\degr$ at the $3\sigma$ level. This plot was
  generated from the solution-specific diameter ranges, but looks the
  same for any other diameter range.  \label{fig:obl_chi2}}
\end{figure}

\begin{figure}
\epsscale{.80}
\plotone{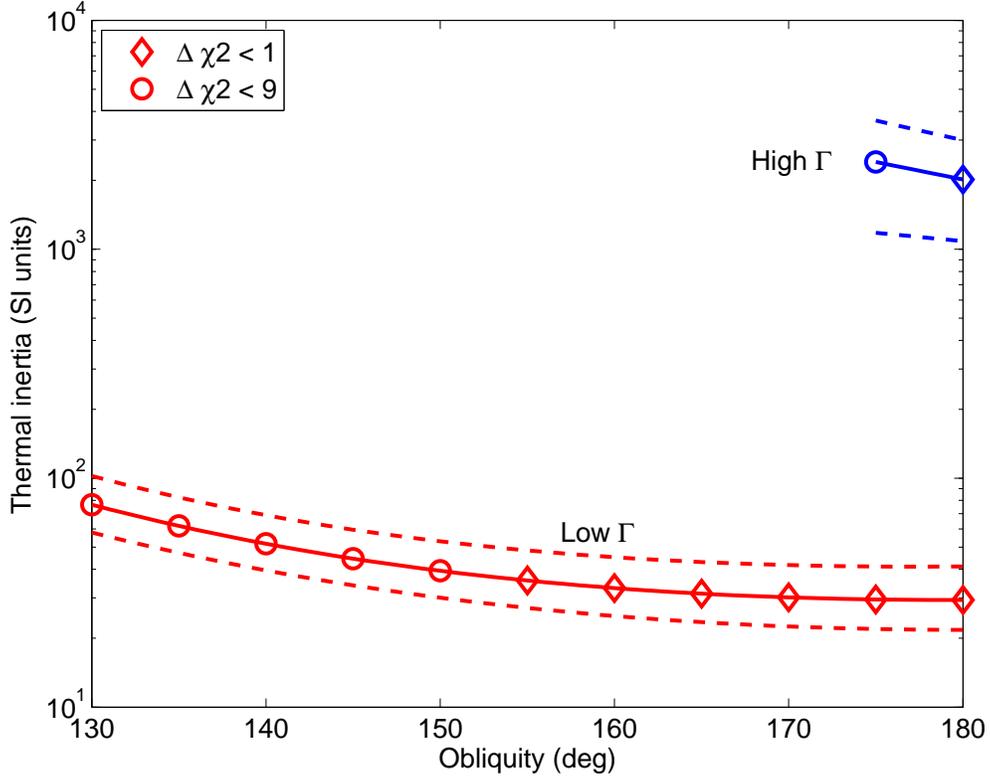}
\caption{Thermal inertia as a function of the obliquity for the
  low-$\Gamma$ and high-$\Gamma$ solutions. Continuous lines give the
  most likely solution based on the numerical simulation; dashed lines
  indicate the $1\sigma$ confidence interval. Symbols indicate $\Delta
  \chi^2$, the difference in $\chi^2$ from the respective minimum
  value of $\chi^2$. Note that $\Delta \chi^2 < 1$ (diamonds) refers
  to the $1\sigma$ and $\Delta \chi^2 < 9$ (circles) to the $3\sigma$
  confidence interval in $\gamma$, as shown in Figure
  \ref{fig:obl_chi2}. It is obvious from this plot that both the
  low-$\Gamma$ and the high-$\Gamma$ solution cover very distinct
  ranges in thermal inertia over the physically meaningful range of
  $\gamma$.  \label{fig:obl_TI}}
\end{figure}

\begin{figure}
\epsscale{.80}
\plotone{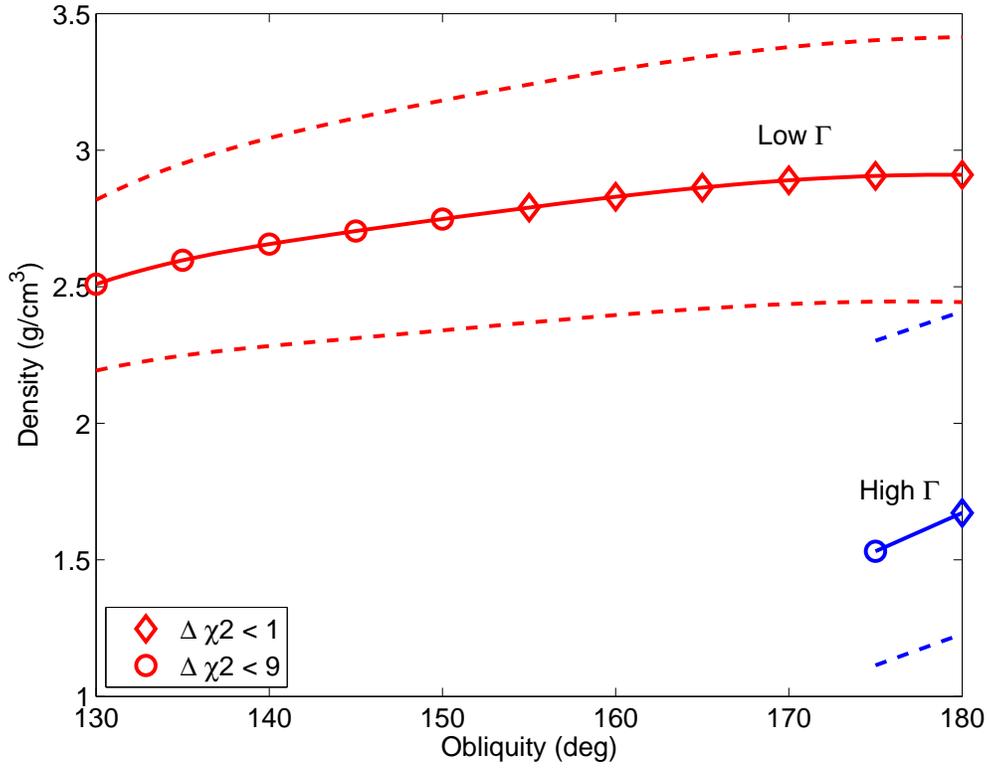}
\caption{Bulk density as a function of the obliquity for the
  low-$\Gamma$ and high-$\Gamma$ solutions. Symbols and lines have the
  same meanings as in Figure \ref{fig:obl_TI}. \label{fig:obl_rho}}
\end{figure}

Based on the solution-specific diameter ranges, we finally constrain
the other physical properties of 2009~BD using the orbital model.
Figure \ref{fig:obl_chi2} shows how obliquity is constrained by the
orbital fit $\chi^2$. By mapping the distribution in obliquity to the
distributions in thermal inertia (Figure \ref{fig:obl_TI}) and bulk
density (Figure \ref{fig:obl_rho}) we obtain our final estimates for
the bulk density, thermal inertia, and total mass for both solutions,
as listed in Table \ref{tbl:solutions}. The reported $1\sigma$ error
bars account for the uncertainties of the input physical parameters
used to model the Yarkovsky accelerations (e.g., diameter and absolute
magnitude) and the uncertainty resulting from the astrometry.
The individual physical properties derived from both the low-$\Gamma$
and the high-$\Gamma$ solution are discussed in Section
\ref{lbl:discussion_properties}.

\begin{deluxetable}{rcc}
\tabletypesize{\scriptsize}
\tablecaption{Physical Properties of 2009~BD\label{tbl:solutions}}
\tablewidth{0pt} 
\tablehead{ 
  \colhead{Parameter} & \colhead{Low-$\Gamma$ Solution} 
  & \colhead{High-$\Gamma$ Solution}}
\startdata 
Diameter $d$ (m) & $2.9\pm0.3$ & $4\pm1$ \\
Albedo $p_V$ & $0.85_{-0.10}^{+0.20}$ & $0.45_{-0.15}^{+0.35}$ \\
Obliquity $\gamma$ (\degr) & $170_{-15}^{+10}$ & $180_{-5}^{+0}$ \\
AMR $\Psi$ ($\times 10^{-4}$ m$^2$~kg$^{-1}$) &$1.8_{-0.2}^{+0.3}$ &  $2.2_{-0.2}^{+0.4}$\\
Bulk Density $\rho$ (g~cm$^{-3}$) & $2.9_{-0.5}^{+0.5}$ & $1.7_{-0.4}^{+0.7}$ \\
Macroporosity (\%) & $10_{-10}^{+20}$ & $45_{-30}^{+15}$ \\
Total Mass (Metric Tons) & $36_{-8}^{+10}$ & $55_{-25}^{+30}$\\
Thermal Inertia $\Gamma$ (SI units) & $30_{-10}^{+20}$ & $2000\pm1000$
\\
\enddata
\tablecomments{Uncertainties depict the $1\sigma$ confidence
  interval.}
\end{deluxetable}


\section{Discussion}
\label{lbl:discussion}

\subsection{Observations and {\it Spitzer} Pointing}
\label{lbl:discussion_observations}

2009~BD was not detected in our observations, and was fainter than
expected (see Figure \ref{fig:sensitivity_finalimage}). Furthermore,
the final sensitivity was lower than predicted: a 5$\sigma$ detection
would have been anticipated at $0.78$~$\mu$Jy in channel 2, whereas we
are only able to derive a $3\sigma$ upper limit at this flux
density. The lower sensitivity is possibly a result of the fact that
2009~BD moved only a short distance over the integration time,
compromising the background removal and increasing the background
noise of the final mosaic. 



To verify that our non-detection is meaningful we have carefully
checked to ensure that the {\it Spitzer} pointing was correct and the
2009~BD ephemeris predictions were accurate. To check the {\it Spitzer}
pointing we have compared the most current predicted position with the
position assumed by {\it Spitzer} for each frame. We find a mean discrepancy
between the {\it Spitzer} pointing and the orbital predictions of
$4.7\pm0.3$\arcsec\ in right ascension and $1.6\pm0.2$\arcsec\ in
declination. This small discrepancy is readily explained by the fact
that the predictions are based on a more recent {\it Spitzer} ephemeris than
was available during the {\it Spitzer} observations; {\it Spitzer} position errors
can be a few hundred km, leading to 3-6\arcsec\ pointing errors for
2009~BD. Therefore, the new position prediction is superior to the
one assumed at the time of the observations. The ellipse in Figure
\ref{fig:sensitivity_finalimage} marks this updated position, which
does not show a detection of 2009~BD.

So far we have shown that the searched position matches the predicted
position, but to have confidence in the non-detection analysis we must
also show that 2009~BD was in fact close to the predicted position. To
this end we compared our nominal plane-of-sky predictions from JPL
Solution 41 to the results of various alternate orbital solutions,
e.g., taking into account $A1$ only, both $A1$ and $A2$, gravitational
effects only, and different outlier rejection schemes
\citep{Carpino2003}, with the result that 2009~BD was always near the
nominal prediction. We also tried to fit synthetic observations that
were ${\sim}5\arcsec$ from the prediction, but this consistently led
to unrealistic residuals for the other observations. From this we
conclude that 2009~BD could not be even ${\sim}5\arcsec$ from its
predicted position.

\subsection{Physical Properties}
\label{lbl:discussion_properties}

By combining the orbital and the thermophysical model we are able to
constrain a variety of physical properties of 2009~BD by taking
advantage of the mutual dependencies of the individual properties. The
range of each physical property is confined based on purely physical
considerations. The only properties that are not derived as part of
the modeling process are $H$, $G$, and $P$ (see Section
\ref{lbl:modeling}), which are based on observations of 2009~BD or
general properties of the asteroid population. We discuss those
physical properties that are constrained in the modeling process.

We derive a volume-equivalent diameter of $2.9\pm0.3$~m and a
geometric albedo of $p_V = 0.85_{-0.10}^{+0.20}$ for the low-$\Gamma$
solution and $d=4\pm1$~m and $0.45_{-0.15}^{+0.35}$ for the
high-$\Gamma$ solution. Both albedo solutions exhibit relatively high
albedos compared to albedo measurements of other NEOs
\citep{Trilling2010, Mainzer2011}. The diameter results agree within
uncertainties, allowing for a more generalized formulation: 2009~BD's
diameter is $2.6 < d < 7$~m at a 3$\sigma$ confidence level. Note that
despite the fact that we did not detect 2009~BD, we are able to
constrain the diameter in both cases with uncertainties that are
rather low compared to the typical 20\% diameter uncertainty derived
from ExploreNEOs data \citep{Harris2011}. We are able to confine the
diameter to such a narrow range due to constraints given by the
physics of the orbital model and the upper-limit flux density derived
from our observations. As a matter of fact, a diameter lower than
2.6~m implies a high Bond albedo, which in turn reduces the size of
the Yarkovsky effect and prevents our model from matching the
magnitude of the observed acceleration. The advantage of this
definition of the lower limit is that we do not {\it apriori} rule out
the possibility that 2009~BD has an extremely high albedo, which is
the case for the low-$\Gamma$ solution. Although relatively rare,
similarly high albedos have been found by both the ExploreNEOs
\citep{Trilling2010} and the NEOWISE projects
\citep{Mainzer2011}. Based on spectral work by \citet{Thomas2011},
both albedo results are compatible with a E/S/V/Q-type taxonomic
classification for 2009~BD.

\begin{figure}
\epsscale{.80}
\plotone{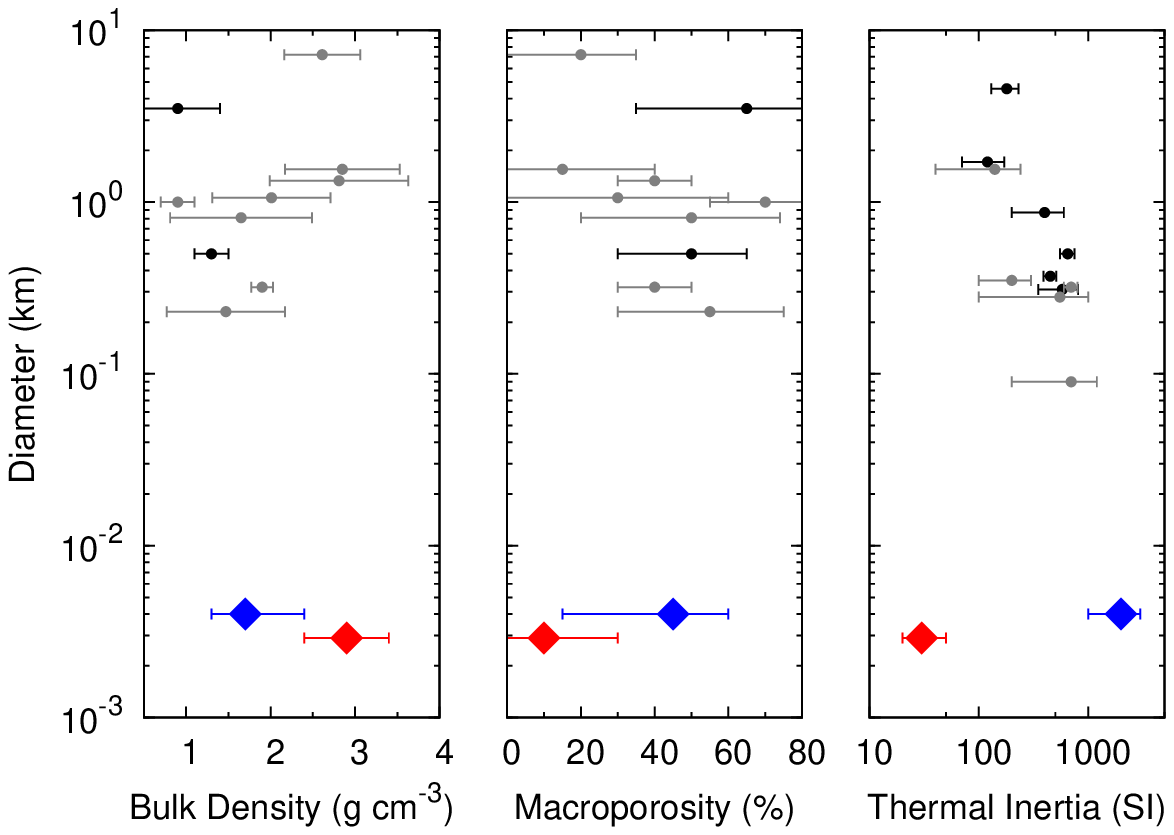}
\caption{Physical properties of known asteroids with diameters of
  10~km or less, as a function of their diameter: bulk density (left),
  macroporosity (center), and thermal inertia (right). Grey circles
  depict stony asteroid types (S/Q/V/E), black circles carbonaceous
  types (B/C) (according to Table \ref{tbl:physical_properties}).  The
  low-$\Gamma$ solution of 2009~BD is indicated as a red diamond, the
  high-$\Gamma$ solution as a blue diamond in each plot. A large
  degree of variety in bulk densities and macroporosity is obvious,
  and there is no clear trend between either of them and diameter.
  Thermal inertia exhibits a slight trend of smaller objects having
  higher thermal inertia \citep[see][]{Delbo2007}.  Data plotted here
  are tabulated in Tables \ref{tbl:solutions} and
  \ref{tbl:physical_properties}.\label{fig:physical_properties}}
\end{figure}

We compare the derived bulk density solutions for 2009~BD with
measured bulk densities of asteroids with diameters of 10~km or
smaller. Density data are taken from the list compiled by
\citet{Baer2012} and the literature (see Table
\ref{tbl:physical_properties}). All measurements are plotted in Figure
\ref{fig:physical_properties}. Macroporosity is derived as unity minus
the ratio of the asteroid's bulk density and the bulk density of
meteorite equivalent material. Meteorite equivalent bulk densities are
taken from \citet{Britt2002}, Table 2. Due to ambiguities in the
identification of the asteroidal origin of meteoritic material, we use
average bulk densities derived from meteoritic material as proxies for
asteroidal bulk densities of individual taxonomic types. Hence, we
assume S and Q-type asteroids to have an average bulk density of
$3.3\pm0.1$~g~cm$^{-3}$, as derived from H/L/LL ordinary chondrites,
and C-type and B-type asteroids to have an average bulk density of
$2.6\pm0.5$~g~cm$^{-3}$, as derived from different types of
carbonaceous chondrites. For V-type asteroids we assume an average
bulk density of $2.9_{-0.4}^{+0.5}$~g~cm$^{-3}$ as derived from
howardite-eucrite-diogenite meteorites
\citep{Macke2011}. \citet{Consolmagno2008} give a mean bulk density of
enstatite chondrites, which likely originate from E-type asteroids, of
$3.5\pm0.2$~g~cm$^{-3}$.

Figure \ref{fig:physical_properties} (center panel) shows that most
objects with diameters 100~m $< d <$ 10~km have macro-porosities
higher than 30\%, consistent with a rubble-pile nature
\citep{Britt2002}. There is no clear trend in either bulk density or
macroporosity with the diameter of the object. Using the derived bulk
density of 2009~BD ($2.9_{-0.5}^{+0.5}$~g~cm$^{-3}$ for the
low-$\Gamma$ solution, $1.7_{-0.4}^{+0.7}$~g~cm$^{-3}$ for the
high-$\Gamma$ solution), and a mean bulk density of
$(3.2\pm0.3)$~g~cm$^{-3}$ derived as the average of the S/Q/V/E-type
asteroid material bulk densities listed above, we derive a degree of
macroporosity of $10_{-10}^{+20}$\% or $45_{-30}^{+15}$\%. Hence, our
low-$\Gamma$ solution is most consistent with a monolithic nature of
2009~BD, whereas the high-$\Gamma$ solution suggests a rubble-pile
nature \citep{Britt2002}.

\begin{deluxetable}{rccccccl}
\tabletypesize{\scriptsize}
\tablecaption{Physical Properties of Small
  Asteroids\label{tbl:physical_properties}} \tablewidth{0pt}
\tablehead{ \colhead{Object} & \colhead{Diameter} & \colhead{Albedo} &
  \colhead{Tax.} & \colhead{Bulk Density} &
  \colhead{MPor.} & \colhead{$\Gamma$} & \colhead{Ref.} \\
  & \colhead{(km)} & & Type & \colhead{(g cm$^{-3}$)} & \colhead{(\%)}
  & \colhead{(SI)} & }
\startdata
(1580)\ Betulia & $4.57\pm0.46$ & $0.08\pm0.02$ & C & ... & ... & $180\pm50$ & 1\\
(1862)\ Apollo & $1.55\pm0.07$ & $0.20\pm0.02$& Q &
$2.85\pm0.68$ & $15_{-15}^{+25}$ & $140_{-100}^{+140}$ & 2, 3\\
(3749)\ Balam & $(7.2\pm0.4)$\tablenotemark{a} &
(0.15)\tablenotemark{a} & (S)\tablenotemark{b} &
($2.61\pm0.45$)\tablenotemark{a} & $20_{-15}^{+15}$& ... & 4 \\
(3908)\ Nyx & $1.0\pm0.2$ & $0.15\pm0.08$ & V & $0.9\pm0.2$\tablenotemark{c} &
$70_{-15}^{+10}$ & ... & 5, 6, 7 \\
(25143)\ Itokawa & $0.320\pm0.001$ & $0.30\pm0.10$ & S & $1.90\pm0.13$
&
$40_{-10}^{+10}$ & $700\pm100$ & 1, 8, 9 \\
(33342)\ 1998~WT24 & $0.35\pm0.04$ & $0.6\pm0.2$ & E & ... & ... & $200\pm100$ & 1, 10\\
(54509)\ YORP & $0.09\pm0.01$ & $0.20\pm0.02$ & S/V & ... & ... & $700\pm500$ & 1 \\
(66391)\ 1999~KW4 & $1.33\pm0.07$ & ($0.25$)\tablenotemark{d}& S &
$2.00\pm0.26$ & $40_{-10}^{+10}$ & ... & 5, 11\\
(101955)\ Bennu & $0.50\pm0.02$ & $0.05\pm0.01$ & B & $1.2\pm0.1$ &
$55_{-15}^{+10}$ & $650\pm100$ & 12, 13\\
(162173)\ 1999~JU3 & $0.87\pm0.03$ & $0.07\pm0.01$ & C & ... & ... &
$400\pm200$ & 3, 14\\
(175706)\ 1996~FG3 & $1.71\pm0.07$ & $0.044\pm0.004$ & C & ... & ... &
$120\pm50$ & 15, 16\\
(185851)\ 2000~DP107 & $0.81\pm0.18$ & ($0.14$)\tablenotemark{d} &
(S)\tablenotemark{d} & $1.65\pm0.84$ & $50_{-30}^{+25}$ & ... & 17\\
(308635)\ 2005~YU55 & $0.31\pm0.01$ & $0.07\pm0.01$ & C & ... & ... &
$580\pm230$ & 18, 19\\
(341843)\ 2008~EV5 & $0.37\pm0.01$ & $0.13\pm0.05$ & C & ... & ... &
$450\pm60$ & 19, 20 \\
2000~UG11 & $0.23\pm0.03$ & (0.23)\tablenotemark{d} &
(S)\tablenotemark{d} & $1.47\pm0.7$ & $55_{-25}^{+20}$ & ... & 21\\
2002~CE26 & $3.50\pm0.40$ & 0.07 & (C)\tablenotemark{e} & $0.9\pm0.5$
& $65_{-30}^{+20}$ & ... & 22\\
2002~NY40 & $0.28\pm0.03$ & $0.34\pm0.06$ & Q & ... & ... & $550\pm450$ & 23, 24 \\
2003~YT1 & $1.06\pm0.06$ & (0.52)\tablenotemark{d} & V & $2.01\pm0.70$
& $30_{-30}^{+30}$ & ... & 25, 26\\
\enddata
\tablecomments{This table lists measured physical properties of
  asteroids with diameters of 10~km or smaller. Data in brackets are
  based on assumptions (see below). In the case of multi-component
  systems, diameters and bulk densities refer to the average numbers
  of the combined system (diameters are those of a volume-equivalent
  sphere). The macroporosity (``MPor.'') is derived through division
  of the asteroid's bulk density by the bulk density of the respective
  meteorite equivalent material (see text for details). Comparison
  data for 2009~BD can be found in Table \ref{tbl:solutions}.}
\tablenotetext{a}{based on an assumed albedo} \tablenotetext{b}{based
  on its Flora family membership \citep{Marchis2008}} \tablenotetext{c}{density estimate assumes a thermal inertia according to \citet{Delbo2007}}
\tablenotetext{d}{albedo derived from the equivalent diameter and the
  $H$ magnitude \citep{SBDB}} \tablenotetext{e}{taxonomic type
  assigned based on albedo determinations by \citet{Thomas2011}}
\tablerefs{ (1) \citet{Mueller2007}; (2) \citet{Rozitis2013}; (3)
  \citet{Bus2002}; (4) \citet{Marchis2008}; (5) \citet{Binzel2004}; 
  (6) \citet{Benner2002}; (7) \citet{Farnocchia2014b}; (8)
  \citet{Fujiwara2006}; (9) \citet{Abe2006}; (10) \citet{Kiselev2002};
  (11) \citet{Ostro2006}; (12) \citet{Mueller2012};
  (13) \citet{Chesley2014}; (14) \citet{Mueller2011b}; (15)
  \citet{Wolters2011}; (16) \citet{Thomas2011}; (17)
  \citet{Margot2002}; (18) \citet{Mueller2013}; (19)
  \citet{Somers2010}; (20) \citet{Ali-Lagoa2013}; (21)
  \citet{Margot2002b}; (22) \citet{Shepard2006}; (23)
  \citet{Roberts2007}; (24) \citet{Mueller2004}; (25)
  \citet{Brooks2006}; (26) \citet{Sanchez2013}.}
\end{deluxetable}


Previous measurements of the thermal inertia of asteroids revealed
values in the range ${\sim}$10--100~SI units for large main belt
asteroids and higher values up to 1000~SI units for NEOs \citep[see,
e.g.,][and references therein]{Delbo2007}. The lower thermal inertia
of large bodies is generally ascribed to the presence of a thick layer
of regolith, which has a low thermal inertia \citep[see Section
\ref{lbl:introduction}, as well as][]{Spencer1989,Putzig2005}. The
right-hand panel of Figure \ref{fig:physical_properties} plots thermal
inertia measurements of asteroids with ${\sim}0.1 < d < 10$~km. The
plot reveals that 2009~BD's thermal inertia is extreme in this range,
irrespective of which of our solutions better describes reality. The
low-$\Gamma$ solution is consistent with the presence of regolith,
whereas the high-$\Gamma$ solution is consistent with a bare-rock
nature of 2009~BD. A slight trend of increasing thermal inertia with
decreasing diameter is visible in Figure
\ref{fig:physical_properties}, which was already discussed by
\citet{Delbo2007}. Our high-$\Gamma$ solution seems to be more
consistent with this trend, presuming that this trend is valid for
asteroids in the size regime of 2009~BD.

\begin{figure}
\epsscale{.80}
\plotone{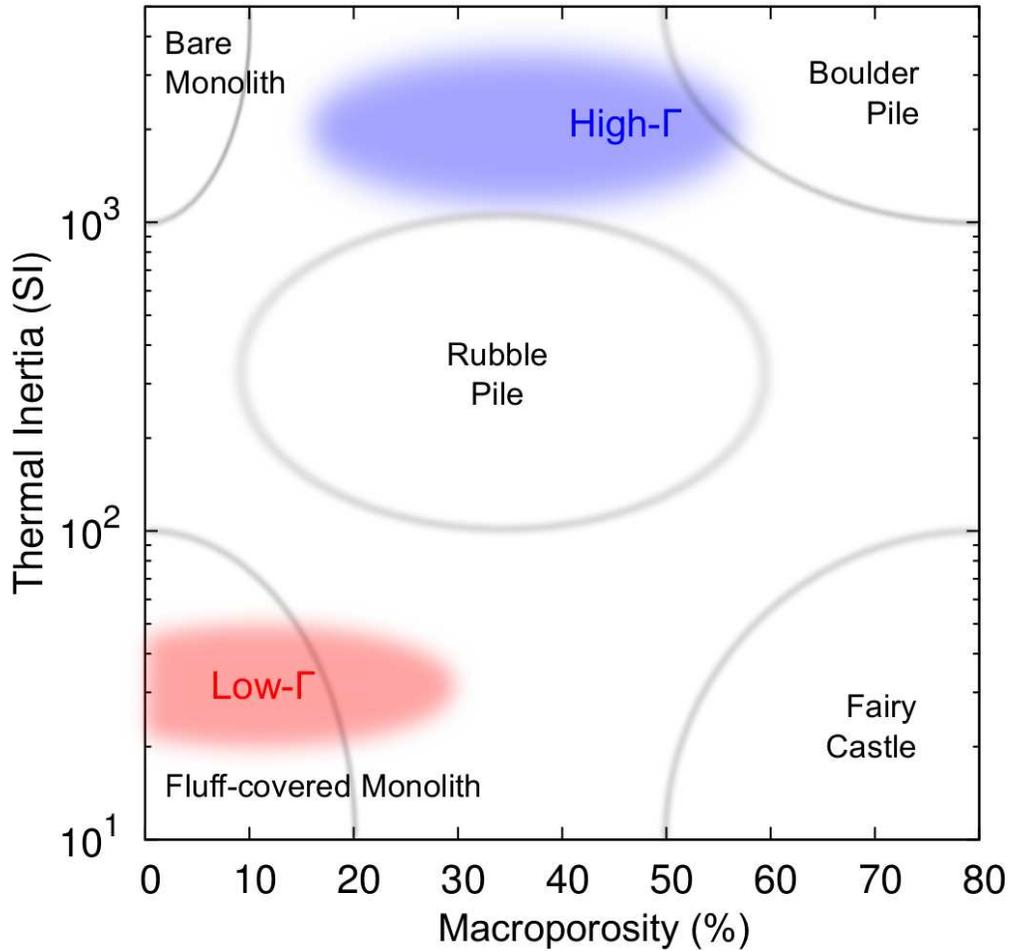}
\caption{Schematic diagram of the nature of 2009~BD. Grey lines
  indicate the different possible configurations as a function of
  thermal inertia and macroporosity. The red and the blue cloud
  symbolize the 1$\sigma$ confidence intervals of the low and
  high-$\Gamma$ solutions, respectively. Note that the border lines of
  the different configurations are not as strict as shown
  here.  \label{fig:schematic}}
\end{figure}

We summarize the properties of 2009~BD created by our two solutions in
Figure \ref{fig:schematic}. The low-$\Gamma$ solution suggests that
2009~BD is a ${\sim}3$~m sized rocky body ($\rho=2.9$~g~cm$^{-3}$)
that is covered with a physically thin but optically thick layer of
regolith-like material, causing it to exhibit a low thermal inertia
($\Gamma=30$~SI units) and a high bulk density. This picture seems
realistic in the sense that models show that even small asteroids can
retain a layer of fine-grained dust on their surfaces
\citep{Scheeres2010,Sanchez2013b}. The picture created by the
high-$\Gamma$ solutions shows 2009~BD as a 4~m-sized rubble-pile
asteroid ($\rho = 1.7$~g~cm$^{-3}$) that consists of individual bare
rock slabs ($\Gamma = 2000$~SI units) and exhibits a macroporosity of
45\%. This scenario seems to be more realistic due to the lower albedo
that is required for this configuration. Despite the fact that some
properties seem to favor one of the two results at a time, we are
unable to rule out either of the configurations based on physical
reasoning. In order to be able to commit to either solution, additional
observations are necessary that are able to pinpoint one decisive
physical property of 2009~BD, e.g., its thermal inertia or its bulk
density. The thermal inertia can be further constrained using
additional infrared observations or using in-situ measurements. The
bulk density can be independently derived by measuring the
gravitational attraction of the object on a nearby body or a
rendezvous spacecraft.

\subsection{Discussion of the Modeling Technique}
\label{lbl:discussion_modeling}

The probabilistic approach taken in this work to constrain the
physical properties of 2009~BD is unique. We have to acknowledge that
a full validation of the methods presented in this paper is not yet
possible. 2009~BD is currently the only asteroid for which both the
Yarkovsky and solar radiation pressure forces can be measured from
astrometric observations, which are used here to constrain the
object's bulk density and thermal inertia.  An independent validation
of our modeling approach would require an asteroid with
non-gravitational perturbations and physical model independently
characterized. The first object for which such a wealth of data is
anticipated will be NEO (101955) Bennu, the OSIRIS-REx mission target.

Instead, we note that both the thermophysical and the dynamical models
are individually well-tested. The thermophysical model used in this
work is based on and has been extensively tested against the model
discussed by \citet{Mueller2007}, which was applied in a number of
publications \citep[e.g.,][]{Harris2007, Mueller2010}. The model of
the Yarkovsky forces is based on work done by
\citet{Vokrouhlicky2000b}, which is used to describe the Yarkovsky
effect observed in a number of objects \citep[e.g.,][]{Chesley2003,
  Vokrouhlicky2008, Farnocchia2013, Chesley2014, Farnocchia2014,
  Farnocchia2014b}. Also, the solar radiation pressure model
\citep{Vokrouhlicky2000} was used to refine orbits of small asteroids
\citep[e.g.,][]{Micheli2012, Micheli2012b, Micheli2013}. Note that in
all previous works in which either model has been used, the resulting
physical properties are within reasonable ranges.

Both the orbital and the thermophysical model assume a spherical shape
of 2009~BD. \citet{Emery2013} have shown that using the real shape of
(101955) Bennu, instead of assuming a spherical shape, lowers the
thermal inertia of that object by a factor of 2. The case of Bennu
shows that shape information can impact the physical parameter
results. We investigate the possible impact of an irregular shape on
our results. 2009~BD has a rotation period $P \geq 3$~hrs
\citep{Tholen2013} with a lightcurve magnitude of ${\geq}0.25$~mag
(B. Ryan, private communication 2014), which suggests an elongation $b/a
\leq 0.8$ for a triaxial ellipsoid with relative dimensions ($a$, $b$,
$c$). Assuming a rotation period $3 < P \ll 25$~hrs, any elongation
effects are averaged out during our 25~hrs integration, leading to
physical properties of a volume-equivalent sphere. We further
investigate the effect of a possible flattening of 2009~BD, by
assuming the shape of a triaxial ellipsoid with axes $b/a = 0.7$ and
$c/b = 0.7$, which is quite typical among larger asteroids. We find
that the smaller cross-section of the triaxial shape compared to that
of a spherical shape requires a reduction of the bulk density of
15--20\% to provide the observed magnitude of the solar radiation
pressure force. A numerical simulation of the Yarkovsky forces
\citep{Vokrouhlicky2000b} suggests that the thermal inertia estimates
might be lower by as much as a factor of 2, which has also been found
for Bennu by \citet{Emery2013}. The smaller thermal inertia in turn
reduces the diameters found for both solutions (see Figures
\ref{fig:d_TI_map} and \ref{fig:diameter_distributions}). Note that
the changes to the individual physical properties found as part of
this simulation are mostly within the uncertainties derived assuming a
spherical shape.  Since there is no certainty on the shape of 2009~BD
we stick to the results based on a simple spherical shape, which are
still valid within the uncertainties, assuming a triaxial
shape. Additional information on the shape of 2009~BD might require a
re-assessment of our results in the future.

\subsection{Implications}

Our results show that the volume-equivalent diameter of 2009~BD,
$2.6 < d < 7.0$~m (3$\sigma$), is most likely smaller than the size
range aimed for in the ARRM mission design (7--10~m, see
above). However, its total mass is roughly 1/10 of the mass aimed for
in the current design, reducing efforts necessary to alter the orbit
of the asteroid. The final decision on 2009~BD's suitability as a
mission target is beyond the scope of this work, but a potential
mission to 2009~BD will be able to resolve the solution degeneracy we
found for the physical properties of this object.

The two scenarios based on our data and presented in Section
\ref{lbl:discussion_properties} show 2009~BD either as a rocky object
covered with regolith-like material or a loose conglomerate of bare
rocks. Either scenario reveals this object as rather exotic compared
to other known asteroids. Hence, 2009~BD may not belong to the normal
population of NEOs that have their origins in the main belt,
accounting for its very Earth-like orbit. It has been suggested that
ejecta from impacts on the Moon could end up in Earth-like orbits;
another possibility is that 2009~BD is a man-made object
\citep{Micheli2012}. A value for $p_V$ of around 0.45 or 0.85 is much
higher than the Moon's albedo albedo of 0.11 \citep{dePater2001},
which would appear to reduce the likelihood that 2009~BD has a lunar
origin. A section of a spent rocket booster would have a high albedo,
but the densities derived here (for both the low and the high-$\Gamma$
solutions) are far higher than values associated with, for example,
hollow rocket fuel tanks. Specifically, the values of the area-to-mass
ratio ($\Psi$) listed in Table \ref{tbl:solutions}, which are
substantially independent of the diameter estimate, are 1--2 orders of
magnitude less than that of artificial objects. The available
astrometry contradicts area-to-mass ratios compatible with a spent
booster.

While our results do not appear to favor any of the more exotic
origins for 2009~BD suggested by its very Earth-like orbit, they
emphasize the puzzling nature of this object and the need for further
observations of this and similar objects.

\section{Summary}

We derive two physically possible solutions for the physical
properties of 2009~BD from our {\it Spitzer} observations, using
thermophysical modeling and modeling of the non-gravitational forces
acting upon this body. The first solution shows 2009~BD as a
$2.9\pm0.3$~m sized massive rock body ($\rho=2.9\pm0.5$~g~cm$^{-3}$)
with an extremely high albedo of $p_V=0.85_{-0.10}^{+0.20}$ that is
covered with regolith-like material, causing it to exhibit a low
thermal inertia ($\Gamma=30_{-10}^{+20}$~SI units). The second
solution suggests 2009~BD to be a $4\pm1$~m-sized rubble-pile asteroid
($\rho = 1.7_{-0.4}^{+0.7}$~g~cm$^{-3}$) with an albedo of
$0.45_{-0.15}^{+0.35}$ that consists of individual bare rock slabs
($\Gamma = 2000\pm1000$~SI units). We are unable to rule out either
solution with the current knowledge.

\acknowledgments

The authors of this work thank Tom Soifer, Director of the {\it
  Spitzer} Space Telescope, for the time allocation to observe
2009~BD. We also would like to thank Paul Chodas for his support and
many informative conversations. We thank an anonymous referee for
useful suggestions that improved this manuscript. D. Farnocchia was
supported for this research by an appointment to the NASA Postdoctoral
Program at the Jet Propulsion Laboratory, California Institute of
Technology, administered by Oak Ridge Associated Universities through
a contract with NASA. The work of S. Chesley was conducted at the Jet
Propulsion Laboratory, California Institute of Technology under a
contract with the National Aeronautics and Space Administration. The
work of D. Vokrouhlick\'{y} was partially supported by the Grant
Agency of the Czech Republic (grant P209-13-01308S). J.~L.\ Hora and
H.~A.\ Smith acknowledge partial support from Jet Propulsion
Laboratory RSA \#1367413. This work is based on observations made with
the {\it Spitzer Space Telescope}, which is operated by the Jet
Propulsion Laboratory, California Institute of Technology under a
contract with NASA.



{\it Facilities:} \facility{Spitzer}.

\clearpage

\end{document}